\documentclass{article}
\usepackage{spconf,amsmath,amssymb,graphicx,booktabs,hyperref,xcolor}

\newcommand{\smoothHours}{100}
\newcommand{\duplexHours}{2,000}
\newcommand{\totalHours}{2,100}

\usepackage{multirow}

\usepackage{pifont}

\newcommand{\cmark}{\textbf{\ding{51}}}
\newcommand{\xmark}{\textbf{\ding{55}}}
\newcommand{\pmark}{\textbf{$\triangle$}}

\title{SmoothConv and DuplexConv: Complementary Mandarin Multi-Party Conversational Speech Corpora for Speech Interaction}
\name{\shortstack{
Chengyou Wang$^{1}$, Mingchen Shao$^{1}$, Chunjiang He$^{1}$, Zeyu Zhu$^{1}$,
Jierui Guo$^{1}$, Bingshen Mu$^{1}$,\\
Zikai Liu$^{1}$, Hanke Xie$^{1}$, Yuhang Dai$^{1,2}$,
Zhou Zhu$^{3}$, Lei Xie$^{1}\sthanks{Corresponding author.}$
}}
\address{
$^{1}$ASLP@NPU,
$^{2}$Soul AI Lab,
$^{3}$QualiaLabs
}
\begin{document}

\maketitle
\fontsize{9}{11}\selectfont
\begin{abstract}
Recent advances in large audio language models (LALMs) have driven the
development of natural and intelligent speech interaction systems. Such
systems need to model complex conversational behaviors, including turn-taking,
overlapping speech, and speaker coordination. Multi-party conversations
provide a realistic setting for studying these behaviors, yet existing
Mandarin conversational corpora often lack synchronized participant-level
speech tracks and comprehensive annotations. In this work, we introduce \textbf{SmoothConv} and \textbf{DuplexConv}, two
complementary Mandarin multi-party conversational speech corpora
totaling 2,100 hours. SmoothConv provides human-verified conversations for reliable
analysis and evaluation, while DuplexConv offers large-scale automatically
annotated conversations through a scalable pipeline for model training. Both
corpora provide synchronized participant-level speech tracks and
multi-dimensional fine-grained annotations. We further release the
\textbf{SmoothConv Benchmark} and evaluate these resources on speech
separation, multi-speaker automatic speech recognition (MSASR), and turn
detection tasks. Experimental results demonstrate the utility of the proposed resources
for multi-party speech interaction modeling. The datasets, benchmark, and related resources are publicly available.\footnote{https://github.com/qualialabsAI/SmoothConv-DuplexConv}
\end{abstract}

\begin{keywords}
Multi-party conversational speech, speech interaction, speech corpus, benchmark
\end{keywords}
\section{Introduction}
\label{sec:intro}

Recent advances in large audio language models (LALMs) have made natural,
real-time, and coherent speech interaction an increasingly important area. Beyond recognizing speech content and speaker identity, interactive speech systems need to understand conversational behaviors, including turn-taking, response timing, and paralinguistic cues. Multi-party conversations provide a natural setting for studying such behaviors, as multiple speakers coordinate their speech through turn exchanges, overlaps, and evolving interaction states. However, existing conversational resources often lack either synchronized participant-level speech tracks or comprehensive annotations of interaction behaviors, limiting their applicability to realistic multi-party speech interaction.

High-quality multi-party conversational data requires two complementary
components: synchronized speech signals from individual participants and
fine-grained annotations describing conversational content, speaker
information, and interaction behaviors. Existing conversational corpora vary in their recording configurations and
annotation schemes.
Switchboard~\cite{godfrey1992switchboard} and Fisher~\cite{cieri2004fisher}
contain natural conversations with separate speaker channels, but are limited
to English two-speaker telephone conversations. CHiME-5~\cite{barker2018chime5}
captures realistic multi-speaker conversations in everyday environments using
multi-channel recordings, but mainly targets robust ASR and speech separation
without comprehensive fine-grained annotations. AMI~\cite{kraaij2005ami} and
AliMeeting~\cite{yu2022m2met} provide multi-party meeting recordings with
speaker annotations, while AISHELL-4~\cite{fu2021aishell4} introduces Mandarin
multi-party meeting data. However, these resources either rely on microphone
array recordings or lack synchronized participant-level speech tracks and
detailed annotations of interaction behaviors. Consequently, publicly available
Mandarin resources remain insufficient for modeling realistic multi-party
conversations.

\begin{table}[t]

\caption{
Comparison with representative conversational corpora.
Multi-Track, Turn, Para., and Attr. denote synchronized participant-level speech
tracks, turn-state annotations, paralinguistic information annotations, and
speaker attributes, respectively.
\cmark/\xmark/\pmark{} indicate full, no, and partial coverage.
}

\label{tab:corpus_comparison}

\centering
\small

\setlength{\tabcolsep}{2pt}

\begin{tabular*}{\columnwidth}{@{\extracolsep{\fill}}lcccccc}

\toprule

Corpora 
& Hours 
& Spk. 
& Multi-Track 
& Turn 
& Para. 
& Attr. \\

\midrule

MagicData-RAMC~\cite{yang2022magicdata}
& 180
& 2
& \xmark
& \xmark
& \xmark
& \xmark \\

Switchboard~\cite{godfrey1992switchboard}
& 260
& 2
& \cmark
& \pmark
& \xmark
& \xmark \\

Fisher~\cite{cieri2004fisher}
& 2,000
& 2
& \cmark
& \xmark
& \xmark
& \xmark \\

CHiME-5~\cite{barker2018chime5}
& 50
& 4
& \pmark
& \xmark
& \xmark
& \xmark
\\
MLC-SLM~\cite{mu2026mlcslm}
& 1,502
& 2
& \xmark
& \xmark
& \xmark
& \xmark
\\
CANDOR~\cite{reece2023candor}
& 850
& 2
& \xmark
& \cmark
& \cmark
& \pmark \\

AMI~\cite{kraaij2005ami}
& 100
& 4
& \cmark
& \cmark
& \pmark
& \xmark \\

AliMeeting~\cite{yu2022m2met}
& 118
& 2--4
& \xmark
& \cmark
& \pmark
& \xmark
\\

AISHELL-4~\cite{fu2021aishell4}
& 120
& 4--8
& \xmark
& \xmark
& \xmark
& \xmark \\

\midrule

\textbf{SmoothConv}
& \smoothHours{}
& 2--4
& \cmark
& \cmark
& \cmark
& \cmark \\

\textbf{DuplexConv}
& \duplexHours{}
& 2--4
& \cmark
& \cmark
& \cmark
& \cmark \\

\bottomrule

\end{tabular*}

\end{table}

Another challenge lies in scaling multi-dimensional annotation to large
conversational datasets. While many conversational corpora provide basic
speech transcriptions and speaker information, fine-grained descriptions of
interaction behaviors and speaker attributes remain limited. Manual annotation
provides reliable and detailed labels, but is costly and difficult to extend
to large-scale data. Automatic annotation improves scalability, yet may
introduce errors in recognition, alignment, and semantic interpretation. This
motivates the development of Mandarin conversational resources that combine
carefully verified data for reliable analysis and automatically constructed
data for scalable model training.
\begin{figure*}[t]
  \centering
  \includegraphics[
  width=\textwidth,
  trim=0 0cm 0.5cm 0cm,
  clip
  ]{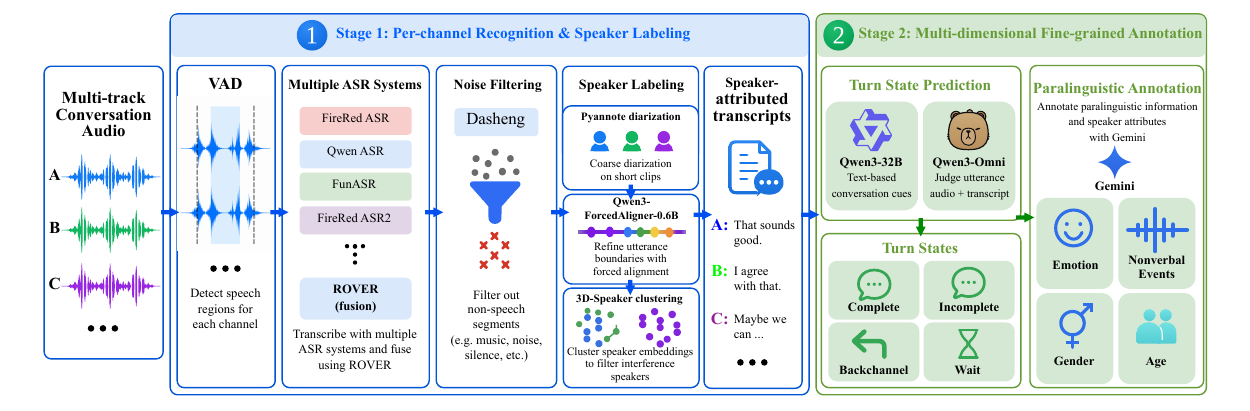}
  \caption{
    Overview of the construction pipeline. Participant-level speech tracks are
    first processed for aligned transcription and speaker labeling, followed by
    multi-dimensional annotation of conversational behaviors and speaker-related
    information.
    }
  \label{fig:pipeline}
\end{figure*}

To address these limitations, we introduce
\textbf{SmoothConv} and \textbf{DuplexConv}, two complementary Mandarin
multi-party conversational speech corpora totaling \totalHours{} hours across
Tutoring and Social Chat scenarios. Both corpora provide synchronized
participant-level speech tracks and multi-dimensional annotations while
serving different research purposes. SmoothConv contains
\smoothHours{} hours of naturally occurring conversations with detailed human
verification, supporting fine-grained analysis and reliable evaluation.
DuplexConv contains \duplexHours{} hours of automatically annotated
conversations generated through a scalable pipeline, supporting large-scale
model training. We further
release the \textbf{SmoothConv Benchmark}, including evaluation data and
protocols for reproducible multi-party speech interaction research. We
evaluate these resources on three representative tasks spanning different
aspects of multi-party speech interaction, including speech separation for
participant-level signal recovery, multi-speaker automatic speech recognition
(MSASR) for speaker-aware content understanding, and turn detection for
conversational behavior modeling. The results demonstrate the utility of
these resources for training and evaluating models on multi-party
conversational speech, highlighting the complementary value of large-scale
automatically annotated data and high-quality human-verified annotations.

\section{Dataset}
\label{sec:dataset}

SmoothConv and DuplexConv are built from synchronized multi-party
conversations collected from real-world online interactions. Unlike scripted,
predefined-topic, or role-playing recordings, these sessions preserve
spontaneous conversational behaviors, including turn exchanges, interruptions,
speech overlaps, and dynamic interaction patterns. All recordings were
collected with informed consent and anonymized before release, with no
personally identifiable information retained.

SmoothConv emphasizes annotation quality through human verification and
supports reliable analysis and evaluation, whereas DuplexConv adopts a
scalable automatic annotation pipeline to provide large-scale training data.

\subsection{Construction Pipeline}
\label{sec:pipeline}

As shown in Fig.~\ref{fig:pipeline}, the pipeline consists of two stages:
per-channel recognition and speaker labeling, followed by multi-dimensional annotation.

\textbf{Per-channel recognition and speaker labeling.} For each participant track, VAD first detects speech regions while preserving
the original timeline synchronization. Multiple ASR systems~\cite{xu2025fireredasr,xu2026fireredasr2s,an2025fun,shi2026qwen3asr} independently
generate transcripts, and ROVER~\cite{fiscus1997rover} fusion combines their
outputs into a unified recognition result. Dasheng~\cite{dinkel2024dasheng} is
then applied to remove non-speech segments. Pyannote diarization~\cite{bredin2023pyannote} and
Qwen3-ForcedAligner-0.6B~\cite{mu2026llmforcedaligner} further refine
speaker attribution and utterance boundaries. Finally, 3D-Speaker
clustering~\cite{chen2024threespeaker} provides consistent speaker labels
across the complete conversation.

\textbf{Multi-dimensional fine-grained annotation.} Based on aligned transcripts and speaker labels, we construct annotations for
conversational turn states, paralinguistic information, and speaker
attributes. To evaluate the quality of automatic annotations, we randomly
sampled a 20-hour subset from DuplexConv for human verification. The sampled
data were independently reviewed by the same annotator team involved in
SmoothConv annotation. A sample is considered valid when the automatically
generated annotations satisfy the predefined quality criteria covering
transcription, speaker labeling, and conversational interaction annotations.
The human verification pass rate on the sampled subset is 82.4\%,
indicating that the proposed automatic annotation pipeline achieves
practical annotation quality for large-scale corpus construction.

For turn-state annotations specifically, each turn is assigned one of four states: \emph{complete}, indicating a
semantically finished utterance; \emph{incomplete}, indicating an unfinished
or interrupted utterance; \emph{backchannel}, representing brief
acknowledgment or feedback; and \emph{wait}, representing a request to pause or
hold the conversation~\cite{easyturn2025,fastturn2026}. Qwen3-32B
~\cite{yang2025qwen3} first predicts candidate turn states from textual
context, and Qwen3-Omni~\cite{xu2025qwen3omni} subsequently incorporates
acoustic information for verification. On the expert-reviewed 20-hour subset,
the automatically generated turn-state labels achieve 91\% agreement with
human annotations. When predictions disagree, the audio-based decision is
preferred for incomplete turns and backchannels, where prosodic cues are
particularly important.

Emotion, nonverbal events, gender, and age attributes are automatically
annotated using Gemini 2.5 Pro~\cite{gemini2023}. Age attributes are
categorized into five groups in DuplexConv: child, teenager, adult, middle-aged, and elderly. Paralinguistic annotations describe emotional
states and nonverbal vocal events, including crying, coughing, throat
clearing, sneezing, sighing, and other vocal behaviors beyond lexical content.
SmoothConv adopts predefined emotion categories to facilitate annotation and
quality control, including neutral, happy, confused, curious, angry, surprise,
sad, contempt, worry, disgust, fear, and sorry. In contrast, DuplexConv uses
free-form emotion descriptions to represent diverse emotional expressions.

\subsection{Dataset Statistics}
\label{sec:released}

The released corpora contain Mandarin conversations from two scenarios:
Tutoring and Social Chat. Each session includes two to four synchronized
participant-level speech tracks, preserving natural speech overlap and
long-range conversational context.

\textbf{SmoothConv} is a human-verified corpus containing 100 hours from
2,503 conversation clips. It covers natural two- to four-speaker
conversations with diverse interaction patterns. Each conversation is
annotated by four annotators and further reviewed by three senior annotators,
covering transcripts, speaker identities, utterance boundaries, turn states,
and interaction attributes. SmoothConv provides reliable supervision for
fine-grained analysis and evaluation of multi-party speech interaction.

DuplexConv contains \duplexHours{} hours from 93,709 clips constructed
through the automatic annotation pipeline. The average clip duration is
76.8\,s, ranging from 8\,s to 618\,s. It includes conversations with two to
four speakers and provides scalable automatically generated annotations to
support large-scale model training.

\subsection{SmoothConv Benchmark}
\label{sec:benchmark}
To facilitate reproducible evaluation, we further release the
\textbf{SmoothConv Benchmark}, a task-oriented evaluation suite derived from
the human-verified SmoothConv corpus. As shown in Table~\ref{tab:benchmark},
the benchmark provides dedicated evaluation subsets for different
multi-party speech interaction tasks.

For speech separation and multi-speaker ASR (MSASR), we construct a shared
evaluation subset containing synchronized participant-level speech tracks.
The subset consists of natural two- to four-speaker conversations and follows
a strict speaker-disjoint partition strategy, ensuring that no speaker
identity appears in both training and evaluation sets. 
Moreover, the average speech overlap ratio is 3.33\%, reflecting the sparse
but challenging overlap patterns in real-world conversations.
\vspace{-3.5mm}
\begin{table}[h]

\caption{
Statistics of the SmoothConv Benchmark subsets.
}
\label{tab:benchmark}

\centering
\small

\setlength{\tabcolsep}{8pt}

\begin{tabular}{lc}

\toprule

\multicolumn{2}{c}{\textbf{Speech Separation / MSASR Evaluation Set}}
\\

\midrule

Duration
& 9.48 h
\\

Clips
& 255
\\

Average duration
& 133.9 s
\\

Speaker configuration
& 2--4 speakers
\\

Scenario
& Tutoring 56.0\% / Social 44.0\%
\\

Speaker split
& Disjoint
\\

Speech overlap
& 3.33\%
\\

\midrule

\multicolumn{2}{c}{\textbf{Turn Detection Evaluation Set}}
\\

\midrule

Turn State
& Samples / Duration (h)
\\

Complete
& 14709 / 9.64
\\

Incomplete
& 3643 / 2.15
\\

Backchannel
& 3080 / 0.42
\\

Wait$^{\dagger}$
& 1000 / 0.71
\\

\bottomrule

\end{tabular}

\vspace{1mm}

{\footnotesize
$^{\dagger}$ Synthesized samples generated only for evaluation.
}

\end{table}
\vspace{-3.5mm}

For turn detection, we construct an independent evaluation subset consisting
of short segmented dialogue clips with turn-state annotations. The subset
mainly contains real-world dialogue segments with \emph{complete},
\emph{incomplete}, and \emph{backchannel} states. Since explicit \emph{wait}
states are relatively sparse in natural conversations, we further introduce
1,000 synthesized \emph{wait} samples. These samples are generated using
DeepSeek V3~\cite{liu2024deepseek} for text generation and IndexTTS2
~\cite{zhou2025indextts2} for speech synthesis. All synthesized samples are
manually checked and used only for evaluation.


















\section{Experiments}
\label{sec:experiments}

We evaluate the proposed corpora on three downstream tasks closely related to
multi-party speech interaction: speech separation, multi-speaker automatic
speech recognition (MSASR), and conversational turn detection. These tasks cover three important aspects of interactive speech processing. Speech separation evaluates whether the corpus can support reliable speaker-level signal recovery, which serves as a prerequisite for subsequent speaker-aware interaction understanding.
MSASR focuses on understanding ``who said what'', providing the fundamental
content representation required for interaction modeling. Turn detection
further investigates conversational dynamics by modeling speaker transitions
and interaction states. 

All training and evaluation splits are strictly separated, with no shared
conversation sessions or speakers. When SmoothConv is used for training, the
corresponding evaluation samples are removed to avoid data leakage.

\begin{table}[ht]
\centering

\caption{
Speech separation results under different training configurations.
SI-SDRi (dB) and CER (\%) are reported.
DC (DuplexConv-derived mixtures), Sim. (simulated mixtures), and SC
(SmoothConv) denote training sources.
In-domain and OOD denote in-domain and out-of-domain evaluations.
}
\label{tab:separation_results}

\small

\setlength{\tabcolsep}{1pt}

\begin{tabular*}{\columnwidth}{
@{\extracolsep{\fill}}lcccc
}

\toprule

\multirow{2}{*}{\centering Training data}
& \multicolumn{2}{c}{In-domain (SI-SDRi$\uparrow$/CER$\downarrow$)}
& \multicolumn{2}{c}{OOD (CER$\downarrow$)}
\\

\cmidrule(lr){2-3}
\cmidrule(lr){4-5}

&
Simulated
& SmoothConv
& AliMeeting
& AISHELL-4
\\

\midrule

Sim.
& \textbf{14.87}/31.99
& 13.55/81.24
& 65.29
& 68.49
\\

DC
& 7.13/37.17
& \textbf{17.26}/\textbf{32.15}
& \textbf{42.26}
& \underline{37.66}
\\

DC+Sim.
& 13.46/\underline{29.29}
& 15.63/45.41
& 53.67
& 40.75
\\

DC+Sim.+SC
& 12.32/29.73
& \underline{15.70}/\underline{42.59}
& \underline{45.48}
& \textbf{36.04}
\\

\bottomrule

\end{tabular*}

\end{table}

\begin{table*}[th]
\centering

\caption{
MSASR results.
DER, cpWER, and tcpWER are reported in percentage (\%).
Baseline denotes the original SoulX-Transcriber model.
DC (DuplexConv) and DC+SC (DuplexConv + SmoothConv) denote adaptation settings.
}
\label{tab:asr_diarization_results}

\small

\setlength{\tabcolsep}{3.5pt}

\begin{tabular}{lccc|ccc|ccc|ccc}

\toprule

\multirow{2}{*}{\centering Training Data}
& \multicolumn{3}{c|}{SmoothConv-Bench}
& \multicolumn{3}{c|}{AliMeeting-Near}
& \multicolumn{3}{c|}{AliMeeting-Far}
& \multicolumn{3}{c}{AMI-SDM}
\\

\cmidrule(lr){2-4}
\cmidrule(lr){5-7}
\cmidrule(lr){8-10}
\cmidrule(lr){11-13}

& DER$\downarrow$ & cpWER$\downarrow$ & tcpWER$\downarrow$
& DER$\downarrow$ & cpWER$\downarrow$ & tcpWER$\downarrow$
& DER$\downarrow$ & cpWER$\downarrow$ & tcpWER$\downarrow$
& DER$\downarrow$ & cpWER$\downarrow$ & tcpWER$\downarrow$
\\

\midrule

Baseline
&40.89&27.04&27.95
&4.48&12.52&11.65
&\textbf{12.82}&\textbf{25.88}&\textbf{25.08}
&16.15&\underline{73.66}&\underline{97.75}
\\

DC
&\underline{13.50}&\underline{17.38}&\underline{18.22}
&\underline{0.71}&\underline{8.27}&\underline{7.56}
&20.29&30.17&30.92
&\underline{15.56}&74.04&97.56
\\

DC+SC
&\textbf{12.73}&\textbf{16.87}&\textbf{17.62}
&\textbf{0.68}&\textbf{8.14}&\textbf{7.43}
&\underline{18.37}&\underline{29.18}&\underline{29.76}
&\textbf{14.57}&\textbf{73.28}&\textbf{97.31}
\\

\bottomrule

\end{tabular}

\end{table*}

\subsection{Speech Separation}
\label{sec:separation}

For speech separation, we train an SPMamba~\cite{li2024spmamba} model using
different training configurations based on simulated mixtures and
DuplexConv-derived mixtures. All settings use 480 hours of training data. The
simulated mixtures are collected from WSJ0-2mix, WHAM!~\cite{Wichern2019WHAM},
WHAMR!~\cite{Maciejewski2020WHAMR}, and Libri2Mix~\cite{cosentino2020librimix},
while DuplexConv mixtures are constructed from participant-level conversational
tracks. Hybrid settings combine the two types of mixtures. Evaluation is performed on simulated mixtures, the SmoothConv Benchmark, and
two out-of-domain conversational datasets, AliMeeting~\cite{yu2022m2met} and
AISHELL-4~\cite{fu2021aishell4}. SI-SDRi and CER are reported, where CER is
computed using the X-ASR-zh-en model on separated outputs. For out-of-domain
datasets, only CER is reported due to the lack of clean source references.

As shown in Table~\ref{tab:separation_results}, simulated and conversational
mixtures exhibit complementary advantages. DuplexConv-only training achieves
competitive performance on conversational evaluation sets, while the hybrid
setting provides a better trade-off between simulated and conversational conditions. These results demonstrate that participant-level speech tracks
from natural conversations can effectively support multi-party speech
separation.

\subsection{Multi-speaker Automatic Speech Recognition}
\label{sec:asr_diarization}

For MSASR, we adapt the SoulX-Transcriber~\cite{dai2026soulx} model using
conversational data from DuplexConv and SmoothConv. LoRA~\cite{hu2022lora}
adaptation is applied while keeping the backbone model unchanged. All models
are fine-tuned for one epoch with a learning rate of $1\times10^{-5}$, a
warmup ratio of 0.05, and LoRA parameters of rank 8 and $\alpha=32$.
All training and evaluation splits are strictly separated. Evaluation is
conducted on the SmoothConv Benchmark and external conversational datasets,
including AliMeeting~\cite{yu2022m2met} and AMI-SDM~\cite{kraaij2005ami}.
DER, cpWER, and tcpWER are reported with a 0.25\,s collar for diarization
evaluation.

As shown in Table~\ref{tab:asr_diarization_results}, DuplexConv adaptation
substantially improves MSASR performance on the SmoothConv Benchmark,
reducing both diarization and transcription errors compared with the baseline.
Combining DuplexConv with SmoothConv further improves performance under
several evaluation conditions, demonstrating the complementary benefits of scalable automatically annotated
data and human-verified annotations. These results indicate that the proposed corpora effectively support
speaker-aware content understanding for multi-party speech interaction. The
performance degradation on far-field AliMeeting further suggests that acoustic
mismatch remains a challenge for cross-domain conversational adaptation.

\begin{table}[t]
\centering

\caption{
Turn detection recall (\%) under different training settings.
ET (EasyTurn training set), DC (DuplexConv), and DC+SC (DuplexConv +
SmoothConv) denote training configurations.
C, I, B, and W denote Complete, Incomplete, Backchannel, and Wait states,
respectively.
}

\label{tab:turn_detection}

\small
\setlength{\tabcolsep}{2.5pt}

\begin{tabular*}{\columnwidth}{
@{\extracolsep{\fill}}lcccccccc
}

\toprule

\multirow{2}{*}{Training}
&
\multicolumn{4}{c}{EasyTurn-Test}
&
\multicolumn{4}{c}{SmoothConv-Bench}
\\

\cmidrule(lr){2-5}
\cmidrule(lr){6-9}

&
C
&
I
&
B
&
W
&
C
&
I
&
B
&
W
\\

\midrule

ET
& 93.33
& 85.00
& \underline{87.00}
& \textbf{98.00}
& 78.29
& \underline{73.03}
& 92.57
& \textbf{95.70}
\\

DC
& \textbf{99.67}
& \underline{87.67}
& 76.00
& 91.00
& \underline{87.67}
& 72.85
& \underline{93.78}
& 89.60
\\

DC+SC
& \underline{98.00}
& \textbf{91.00}
& \textbf{93.00}
& \underline{96.00}
& \textbf{89.20}
& \textbf{74.66}
& \textbf{94.15}
& \underline{91.30}
\\
\bottomrule

\end{tabular*}

\end{table}

\subsection{Turn Detection}
\label{sec:turn_detection}

Turn detection evaluates conversational behavior modeling, including speaker
transitions and interaction states. We train FastTurn~\cite{fastturn2026} using three training configurations: EasyTurn~\cite{easyturn2025}, DuplexConv
(DC), and DuplexConv combined with SmoothConv (DC+SC). Recall is reported for
four interaction states: \emph{complete}, \emph{incomplete}, \emph{backchannel}, and \emph{wait}.

As shown in Table~\ref{tab:turn_detection}, different datasets provide
complementary supervision for turn-state modeling. On EasyTurn-Test, the
EasyTurn-trained model achieves the best performance on the \emph{wait} state due to
its dedicated annotations, while DuplexConv achieves stronger complete-turn
recognition with large-scale conversational training data. Combining
DuplexConv with SmoothConv further improves incomplete and backchannel
detection, indicating that high-quality annotations help capture subtle
interaction boundaries.

On the SmoothConv Benchmark, DC+SC achieves the best performance on most
states. The improvements on incomplete and backchannel states demonstrate the
value of combining large-scale conversational data with reliable annotations
for modeling speaker coordination. The \emph{wait} state remains challenging because
explicit waiting behaviors are relatively rare in natural conversations, and
the evaluation set contains synthesized samples, resulting in a potential gap
from real interaction patterns.
















\section{Conclusion}
\label{sec:conclusion}
In this work, we introduce SmoothConv and DuplexConv, two
complementary Mandarin multi-party conversational corpora with
participant-level tracks and multi-dimensional annotations. SmoothConv
provides high-quality human-verified data for analysis and evaluation,
while DuplexConv enables large-scale model training with automatically
annotated conversations. We further release the SmoothConv Benchmark, a
task-oriented evaluation suite covering different aspects of multi-party
speech interaction. Experiments on speech separation, MSASR, and turn
detection demonstrate the utility of these publicly available resources
for modeling natural multi-party speech interaction.

\bibliographystyle{IEEEbib}
\bibliography{refs}

@inproceedings{yang2022magicdata,
  title     = {Open Source {MagicData-RAMC}: A Rich Annotated {Mandarin} Conversational {(RAMC)} Speech Dataset},
  author    = {Yang, Zehui and others},
  booktitle = {Proc. Interspeech},
  pages     = {1736--1740},
  year      = {2022},
  doi       = {10.21437/Interspeech.2022-729}
}

@inproceedings{kraaij2005ami,
  title     = {The {AMI} Meeting Corpus: A Pre-announcement},
  author    = {Carletta, Jean and others},
  booktitle = {Proc. MLMI},
  series    = {LNCS},
  volume    = {3869},
  pages     = {28--39},
  year      = {2005},
  publisher = {Springer},
  doi       = {10.1007/11677482_3}
}

@article{dai2026soulx,
  title         = {{SoulX-Transcriber}: A Robust End-to-End Framework for Multi-Speaker Speech Transcription},
  author        = {Dai, Yuhang and others},
  journal       = {arXiv preprint arXiv:2606.02400},
  year          = {2026},
  eprint        = {2606.02400},
  archivePrefix = {arXiv},
  primaryClass  = {eess.AS},
  doi           = {10.48550/arXiv.2606.02400}
}

@inproceedings{godfrey1992switchboard,
  title     = {{SWITCHBOARD}: Telephone Speech Corpus for Research and Development},
  author    = {Godfrey, John J. and Holliman, Edward C. and McDaniel, Jane},
  booktitle = {Proc. ICASSP},
  volume    = {1},
  pages     = {517--520},
  year      = {1992},
  doi       = {10.1109/ICASSP.1992.225858}
}

@inproceedings{cieri2004fisher,
  title     = {The {Fisher} Corpus: A Resource for the Next Generations of Speech-to-Text},
  author    = {Cieri, Christopher and Miller, David and Walker, Kevin},
  booktitle = {Proc. LREC},
  pages     = {69--71},
  year      = {2004}
}

@article{reece2023candor,
  title   = {The {CANDOR} Corpus: Insights from a Large Multimodal Dataset of Naturalistic Conversation},
  author  = {Reece, Andrew and others},
  journal = {Science Advances},
  volume  = {9},
  number  = {13},
  pages   = {eadf3197},
  year    = {2023},
  doi     = {10.1126/sciadv.adf3197}
}

@inproceedings{fu2021aishell4,
  title     = {{AISHELL-4}: An Open Source Dataset for Speech Enhancement, Separation, Recognition and Speaker Diarization in Conference Scenario},
  author    = {Fu, Yihui and others},
  booktitle = {Proc. Interspeech},
  pages     = {3665--3669},
  year      = {2021},
  doi       = {10.21437/Interspeech.2021-1397}
}

@inproceedings{easyturn2025,
  title     = {{Easy Turn}: Integrating Acoustic and Linguistic Modalities for Robust Turn-Taking in Full-Duplex Spoken Dialogue Systems},
  author    = {Li, Guojian and others},
  booktitle = {Proc. ICASSP},
  pages     = {16957--16961},
  year      = {2026},
  doi       = {10.1109/ICASSP55912.2026.11463929}
}

@article{fastturn2026,
  title         = {{FastTurn}: Unifying Acoustic and Streaming Semantic Cues for Low-Latency and Robust Turn Detection},
  author        = {Wang, Chengyou and others},
  journal       = {arXiv preprint arXiv:2604.01897},
  year          = {2026},
  eprint        = {2604.01897},
  archivePrefix = {arXiv},
  primaryClass  = {cs.SD},
  doi           = {10.48550/arXiv.2604.01897}
}

@article{li2024spmamba,
  title         = {{SPMamba}: State-space Model is All You Need in Speech Separation},
  author        = {Li, Kai and Chen, Guo and Yang, Runxuan and Hu, Xiaolin},
  journal       = {arXiv preprint arXiv:2404.02063},
  year          = {2024},
  eprint        = {2404.02063},
  archivePrefix = {arXiv},
  primaryClass  = {cs.SD},
  doi           = {10.48550/arXiv.2404.02063}
}

@article{yang2025qwen3,
  title         = {{Qwen3} Technical Report},
  author        = {{Qwen Team}},
  journal       = {arXiv preprint arXiv:2505.09388},
  year          = {2025},
  eprint        = {2505.09388},
  archivePrefix = {arXiv},
  primaryClass  = {cs.CL},
  doi           = {10.48550/arXiv.2505.09388}
}

@article{xu2025qwen3omni,
  title         = {{Qwen3-Omni} Technical Report},
  author        = {{Qwen Team}},
  journal       = {arXiv preprint arXiv:2509.17765},
  year          = {2025},
  eprint        = {2509.17765},
  archivePrefix = {arXiv},
  primaryClass  = {cs.CL},
  doi           = {10.48550/arXiv.2509.17765}
}

@article{shi2026qwen3asr,
  title         = {{Qwen3-ASR} Technical Report},
  author        = {Shi, Xian and others},
  journal       = {arXiv preprint arXiv:2601.21337},
  year          = {2026},
  eprint        = {2601.21337},
  archivePrefix = {arXiv},
  primaryClass  = {cs.CL},
  doi           = {10.48550/arXiv.2601.21337}
}

@article{gemini2023,
  author       = {{Gemini Team}},
  title        = {{Gemini}: {A} Family of Highly Capable Multimodal Models},
  journal      = {CoRR},
  volume       = {abs/2312.11805},
  year         = {2023},
  doi          = {10.48550/arXiv.2312.11805},
  eprinttype   = {arXiv},
  eprint       = {2312.11805},
}

@inproceedings{fiscus1997rover,
  title     = {A Post-Processing System to Yield Reduced Word Error Rates: Recognizer Output Voting Error Reduction ({ROVER})},
  author    = {Fiscus, Jonathan G.},
  booktitle = {Proc. ASRU},
  pages     = {347--354},
  year      = {1997},
  doi       = {10.1109/ASRU.1997.659110}
}

@inproceedings{dinkel2024dasheng,
  title     = {Scaling Up Masked Audio Encoder Learning for General Audio Classification},
  author    = {Dinkel, Heinrich and Yan, Zhiyong and Wang, Yongqing and Zhang, Junbo and Wang, Yujun and Wang, Bin},
  booktitle = {Proc. Interspeech},
  pages     = {547--551},
  year      = {2024},
  doi       = {10.21437/Interspeech.2024-246}
}

@inproceedings{bredin2023pyannote,
  title     = {pyannote.audio 2.1 Speaker Diarization Pipeline: Principle, Benchmark, and Recipe},
  author    = {Bredin, Herv{\'e}},
  booktitle = {Proc. Interspeech},
  pages     = {1983--1987},
  year      = {2023},
  doi       = {10.21437/Interspeech.2023-105}
}

@inproceedings{chen2024threespeaker,
  title     = {{3D-Speaker-Toolkit}: An Open-Source Toolkit for Multimodal Speaker Verification and Diarization},
  author    = {Chen, Yafeng and others},
  booktitle = {Proc. ICASSP},
  pages     = {1--5},
  year      = {2025},
  doi       = {10.1109/ICASSP49660.2025.10888389}
}

@inproceedings{yu2022m2met,
  title     = {{M2MeT}: The {ICASSP} 2022 Multi-Channel Multi-Party Meeting Transcription Challenge},
  author    = {Yu, Fan and others},
  booktitle = {Proc. ICASSP},
  pages     = {6167--6171},
  year      = {2022},
  doi       = {10.1109/ICASSP43922.2022.9746465}
}

@inproceedings{hu2022lora,
  title     = {{LoRA}: Low-Rank Adaptation of Large Language Models},
  author    = {Hu, Edward J. and others},
  booktitle = {Proc. ICLR},
  year      = {2022}
}

@article{xu2025fireredasr,
  title={Fireredasr: Open-source industrial-grade mandarin speech recognition models from encoder-decoder to llm integration},
  author={Xu, Kai-Tuo and Xie, Feng-Long and Tang, Xu and Hu, Yao},
  journal={arXiv preprint arXiv:2501.14350},
  year={2025}
}

@article{xu2026fireredasr2s,
  title={Fireredasr2s: A state-of-the-art industrial-grade all-in-one automatic speech recognition system},
  author={Xu, Kaituo and others},
  journal={arXiv preprint arXiv:2603.10420},
  year={2026}
}

@article{an2025fun,
  title={Fun-asr technical report},
  author={An, Keyu and others},
  journal={arXiv preprint arXiv:2509.12508},
  year={2025}
}

@article{cosentino2020librimix,
  title         = {{LibriMix}: An Open-Source Dataset for Generalizable Speech Separation},
  author        = {Cosentino, Joris and others},
  journal       = {arXiv preprint arXiv:2005.11262},
  year          = {2020},
  eprint        = {2005.11262},
  archivePrefix = {arXiv},
  primaryClass  = {eess.AS},
  doi           = {10.48550/arXiv.2005.11262}
}

@inproceedings{wichern2019wham,
  title     = {{WHAM!}: Extending Speech Separation to Noisy Environments},
  author    = {Wichern, Gordon and others},
  booktitle = {Proc. Interspeech},
  pages     = {1368--1372},
  year      = {2019},
  doi       = {10.21437/Interspeech.2019-2821}
}

@inproceedings{maciejewski2020whamr,
  title     = {{WHAMR!}: Noisy and Reverberant Single-Channel Speech Separation},
  author    = {Maciejewski, Matthew and others},
  booktitle = {Proc. ICASSP},
  pages     = {696--700},
  year      = {2020},
  doi       = {10.1109/ICASSP40776.2020.9053327}
}

@article{liu2024deepseek,
  title     = {{DeepSeek-V3 Technical Report}},
  author    = {Liu, Aixin and Feng, Bei and Xue, Bing and Wang, Bingxuan
               and Wu, Bochao and Lu, Chengda and Zhao, Chenggang
               and Deng, Chengqi and Zhang, Chenyu and Ruan, Chong
               and others},
  journal   = {arXiv preprint arXiv:2412.19437},
  year      = {2024}
}

@article{zhou2025indextts2,
  title     = {{IndexTTS2}: A Breakthrough in Emotionally Expressive and
               Duration-Controlled Auto-Regressive Zero-Shot Text-to-Speech},
  author    = {Zhou, Siyi and Zhou, Yiquan and He, Yi and Zhou, Xun
               and Wang, Jinchao and Deng, Wei and Shu, Jingchen
               and others},
  journal   = {arXiv preprint arXiv:2506.21619},
  year      = {2025}
}

@inproceedings{barker2018chime5,
  title     = {The fifth '{CHiME}' Speech Separation and Recognition Challenge:
               Dataset, Task and Baselines},
  author    = {Barker, Jon and Watanabe, Shinji and Vincent, Emmanuel
               and Trmal, Jan},
  booktitle = {Proc. Interspeech},
  pages     = {1561--1565},
  year      = {2018},
  doi       = {10.21437/Interspeech.2018-1768}
}

@inproceedings{mu2026mlcslm,
  title     = {{Summary on the Multilingual Conversational Speech Language
               Model Challenge: Datasets, Tasks, Baselines, and Methods}},
  author    = {Mu, Bingshen and Guo, Pengcheng and Sun, Zhaokai
               and Wang, Shuai and Liu, Hexin and others},
  booktitle = {Proc. ICASSP},
  year      = {2026}
}

@inproceedings{mu2026llmforcedaligner,
  title     = {{LLM-ForcedAligner}: A Non-Autoregressive and Accurate {LLM-Based} Forced
               Aligner for Multilingual and Long-Form Speech},
  author    = {Mu, Bingshen and others},
  booktitle = {Proc. ACL},
  pages     = {25627--25638},
  year      = {2026},
  doi       = {10.18653/V1/2026.ACL-LONG.1174}
}

\end{document}